\documentclass[runningheads,8pt]{llncs}
\pdfoutput=1

\usepackage{epsfig,endnotes,listings}
\usepackage{pifont}
\usepackage{graphicx}
\usepackage{caption}
\usepackage{booktabs}
\usepackage{xcolor}
\usepackage{placeins}
\usepackage{hanging}

\newcommand{\tablestyle}{\sffamily\footnotesize\centering}

\title{Not quite a piece of CHERI-cake: Are new digital security by design architectures usable?}
\titlerunning{Not a piece of CHERI-cake}
\author{Maysara Alhindi \and Joseph Hallett}
\institute{Bristol Cyber Security Group; University of Bristol}
\authorrunning{Alhindi and Hallett}

\newcommand{\lstbg}[3][0pt]{{\fboxsep#1\colorbox{#2}{\strut #3}}}

\lstdefinelanguage{diff}{
  basicstyle=\ttfamily\small,
  morecomment=[f][\lstbg{red!20}]-,
  morecomment=[f][\lstbg{green!20}]+,
  morecomment=[f][\textit]{@@},
  tabsize=4,
  keepspaces,
  breaklines=true,
  breakatwhitespace=false, 
  frame=single,
  showspaces=false,
  showstringspaces=false
}

\newcommand{\code}[3]{
    \lstinputlisting[
        caption=#1,
        label={lst:#1},
        language=#2,
        style=codeStyle,
    ]{#3}
}

\definecolor{codegreen}{rgb}{0,0.6,0}
\definecolor{codegray}{rgb}{0.5,0.5,0.5}
\definecolor{codepurple}{rgb}{0.58,0,0.82}
\definecolor{backcolour}{rgb}{0.95,0.95,0.92}

\lstdefinestyle{codeStyle}{
    commentstyle=\color{codegray},
    keywordstyle=\color{blue},
    numberstyle=\tiny\color{codegray},
    stringstyle=\color{codegreen},
    basicstyle=\ttfamily\footnotesize,
    breakatwhitespace=false,         
    breaklines=true,                 
    keepspaces=true,                 
    numbers=left,       
    numbersep=5pt,                  
    showspaces=false,                
    showstringspaces=false,
    showtabs=false,                  
    tabsize=2,
}

\begin{document}
\maketitle

\begin{abstract}
  A digital security-by-design computer architecture, like CHERI, lets you program without fear of buffer overflows or other memory safety errors, but CHERI also rewrites some of the assumptions about how C works and how fundamental types (such as pointers) are implemented in hardware. We conducted a usability study to examine how developers react to the changes required by CHERI when porting software to run on it. We find that developers struggle with CHERI's display of warnings and errors and a lack of diverse documentation.
\end{abstract}

\section{Introduction}

Memory correctness errors have been a plague on software development since the dawn of programming.  Programming languages, like C and C++, allow programmers to write code that manipulates pointers with little to no checks. Whilst this can allow for some very fast code; it also can lead to bugs, errors and in the worst case the complete compromise of the computer and arbitrary code execution. CHERI~\cite{watson2015cheri} is a novel computer architecture that aims to fix this. Rather than ensuring that the programs do not contain bugs (an approach taken by static analysis tools and safe-programming languages like Haskell or Rust) CHERI changes how pointers are implemented at a hardware level, including bounds and usage information that means that memory correctness errors cause hardware exceptions \emph{before} they can do any harm.  Computers using this new architecture are, slowly, starting to become available to researchers and developers through Arm's Morello architecture~\cite{arm-morello}.

CHERI promises that in order for developers to port their software to be run on CHERI-supported hardware, they only need to implement minimal changes. But if a program is doing pointer arithmetic or switching between integers and pointers, as is sometimes done in C, then changes can become necessary, and can be surprising to low-level programmers. An example of changes developers have to undertake is shown in Listing \ref{cheri-change} \cite{watson2020cheri}. Developers do not have to set the capability bound or permissions, as all of such details are inferred by the compiler, but they have to use correct data types and adhere to some patterns.

\begin{lstlisting}[language=diff,caption=This change is made to use a provenance-preserving type (a type that preserves the pointers' capability metadata). This example is provided in CHERI's introduction guide.,label=cheri-change]
- char *example_bad(long ptr_or_int) {
  return strdup((const char *)ptr_or_int);
}
+ char *example_good(intptr_t ptr_or_int) {
  return strdup((const char *)ptr_or_int);
}
\end{lstlisting}

This paper explores what happens when developers are faced with a new architecture like CHERI. How do they approach fixing problems and working with code when their mental models and assumptions about how programming languages work are misaligned?  What \emph{usability smells}~\cite{patnaik2019usability} (indicators that usability issues may be present) appear when developers are struggling to debug CHERI specific issues.

Specifically, this study answers the following research questions:

\begin{description}
  \item[RQ 1.]: How do developers deal with CHERI-specific warnings and errors?
  \item[RQ 2.]: What are the current pain points (usability issues) for developers who are new to CHERI? and what can be done to help them?
\end{description}

Our findings are as follows:
\begin{itemize}
  \item Developers struggled with CHERI's display of warnings and errors, and found it challenging to apply their understanding of CHERI's abstractions to fix the code.
  \item We identified a new usability smell (floundering) to capture when developers try solutions at random in the hopes that \emph{something} will make the issue go away.
  \item Developers are cautious when approaching a new platform, and small differences can confuse them.  Documentation is consulted but developers struggled to find useful help unless the errors are stated explicitly and searchable.
  \item CHERI lacks a targeted, end-developer focused, documentation and tutorials, which leads to developers becoming frustrated.
\end{itemize}

\section{Background and Related Work}

The security model of CHERI is inspired by a substantial amount of early research that investigated different security models for computer systems~\cite{anderson1972computer}. Based on their work on Multics operating system, Saltzer and Schroeder surveyed different security mechanisms that can be applied to operating systems, and they discussed the concepts of access controls and capabilities~\cite{Saltzer_1974}.

CHERI implements capability-based security in hardware by associating capability metadata to pointers. Capability-based security is not a new concept, as many different systems were designed to implement such a model. At their core, capabilities represent unforgeable tokens of authority, combining both the identification of a resource and the permissions to perform operations on it. 

Capability-based models decentralize control and focus on the possession of capabilities as the determinant of access, if a process holds a capability, it can use the associated resource with the specified rights. This model of access makes delegation straightforward, as capabilities can be passed directly between entities without requiring additional permission checks, it also avoids falling into the confused deputy problem, where a process is tricked into using its higher-level privileges to perform actions on behalf of a malicious user \cite{miller2003capability}.

There is a wide spectrum of solutions that implement capability-based security, ranging from sandboxing mechanisms like Capsicum to programming languages like E, and full operating systems such as Hydra \cite{wulf1974hydra,richardson1993design}. Fuchsia is a more recent example of operating systems that adopt a capability-based security model \cite{fuchsia_code}. Fuchsia adopts a capability-based model by representing all system resources as objects, where the security model is based on the possession of unforgeable handles to~objects.

\subsection{Usable Security}

Several researches highlighted the usability issues associated with security software~\cite{maass2016systematic,motiee2010windows,schreuders2013state}. When it comes to security APIs, there is a lot of work around the usability of crypto APIs, such as the work by Acar et al. that compares the usability of different crypto APIs, and the work by Mindermann et al. that focuses on crypto APIs in Rust.

There are other studies that investigate the usability of sandboxing and access control mechanisms. Schreuders et al conducted a usability study to assess the usability of SELinux, AppArmor, and their own solution FBAC-LSM. The researchers found out that SELinux and AppArmor are not designed with usability in mind~\cite{schreuders2012towards}. Alhindi et al. conducted a usability study on Seccomp and found that developers approach implementing a sandbox in completely different ways, and that sometimes, the usabiliy issues associated with Seccomp lead to developers implementing an ineffective sandbox~\cite{alhindi2025playing}.

CHERI platform promises to improve the security of programs and prevent a wide range of memory vulnerabilities to be exploited. Most of the time, developers do not need to make many changes to run on CHERI-supported hardware. We find that whilst the changes required are rare, they do cause significant confusion.

\subsection{Usability Smells}

Developers struggle to use security and crypto APIs. Green~and~Smith suggested 10 principles to help API developers ensure that their APIs are both usable and secure~\cite{green2016developers}. Patnaik~et~al{.} took these principles and looked for signs that they were being violated in over 2,000 Stack Overflow posts~\cite{patnaik2019usability}. Taking inspiration from Fowler's \emph{code smells}~\cite{fowler2018refactoring}, they identified 16 \emph{usability smells} organised into 4 \emph{whiffs} to describe when APIs may have an issue with usability; and are summarised in Table~\ref{tab:patnaik}. We build on Patnaik~et~al{.} work by utilizing these usability whiffs in our analysis methodology. 

\begin{table}
  \tablestyle
  \begin{tabular}{l l}
    \toprule
    Whiff & Description\\
    \midrule
    Needs a super sleuth & The documentation is hard to figure out.\\
    Confusion reigns & The API is confusing to developers.\\
    Needs a post-mortem & It is hard to debug what went wrong.\\
    Doesn't play well with others & The API is hard to integrate with other tech.\\
    \bottomrule\\
  \end{tabular}
  \caption{Summary of Patnaik's four code whiffs indicating the possible presence of usability issues.}
  \label{tab:patnaik}
\end{table}

\FloatBarrier{}

\section{Methodology}

\subsection{Recruitment}

Participants were recruited through professional networks and through direct advertisement to students who had taken courses in C programming, none of our participants had prior experience with porting programs to CHERI. Prior to the study, participants were asked to read two technical reports introducing them to CHERI's architecture: \emph{An Introduction to CHERI}~\cite{watson2019introduction} and \emph{CHERI C/C++ Programming Guide}~\cite{watson2020cheri}. These reports are the recommended readings for developers new to CHERI~\cite{cheri_doc}. The first document gives a general introduction to CHERI architecture, capabilities, and the theory behind it, while the second document focuses on C/C++ CHERI-related APIs on how to manage and check capabilities and it also talks about changes developers have to undertake when porting their programs to CHERI and the warnings they might encounter. These documents are long, 43 and 33 pages, and full of technical details, however, we could not find an easier resource that would explain the needed details. Participants were paid \pounds{}50 in vouchers for completing the study. In total nine participants were recruited to the~study.

At the beginning of the sessions, we asked participants the following two questions to ensure they read the documents and understand the basics of CHERI:

\begin{enumerate}
    \item Can you describe what a CHERI capability is?
    \item What changes do developers have to make to port their programs to CHERI?
\end{enumerate}

Participants who had not read the documentation or who could not answer the above questions were rejected from the study (four participants were rejected as such). Additionally, we asked the participants the same questions at the end of the study to look for learning effects.

\subsection{Study environment}

The study sessions were held in person and were two hours, participants were given access to a laptop that has a QEMU-based CHERI-BSD virtual machine installed on it and many code editors such as Sublime, VScode and Vim, as the participants will be solving a coding task. We also developed a shell script to compile, copy, and run participants' solutions on the CHERI-BSD machine. We thought it would be easier for developers to write code in the host system as they would have more editors available, better performance, and they do not have to worry about how to compile binaries to CHERI, making the participants focus only on solving the task. Before the session, we introduced the participants to the study environment and showed them how to compile and test their code. We also asked the participants to sign a consent form. The audio of the sessions was recorded and later transcribed and anonymized by removing participants’ names and identifiable information. The study was approved by the university's ethics committee, and anonymized transcripts of the sessions are available as open~access \footnote{https://github.com/uob-jh/not-a-piece-of-cheri-cake}.

\subsection{Tasks}

To analyze developers' approaches to working with CHERI, we set them two programming tasks: porting an application to CHERI and reviewing a series of pull requests for CHERI systems. Developers were asked to narrate their thought processes and express any struggles or issues they faced. The developers were given one hour for each task and tasks were performed in a random order to minimize learning effects. Developers had access to the internet and any other documentation they felt they needed throughout the task. The interviews were semi-structured, while we followed a similar protocol throughout the interviews, we asked clarifying questions if participants exhibited interesting behavior. Prior to the actual interview sessions, we carried out three pilot sessions and modified the task and protocol according to the feedback we received.

\subsubsection*{Porting a program to CHERI}

The porting task aims to explore how developers react in the face of CHERI-specific programming errors and warnings; whether they understand what they mean and how they are related to the underlying architecture, and whether they can recall information from CHERI's documentation to correctly fix the code. 

Developers were given a C implementation of a program similar to \texttt{cat}, the program processes a file, outputs its content, and writes text to it. The C program is simple, it is only 89 lines of code, and it performs its functionality using standard libc calls. The program holds similarities to one developed in CHERI exercises \cite{Watson_Davis_2020}. The code is included in the Appendix. Developers were asked to port the program using the capability mode (meaning that it supports CHERI hardware) and run it on the CHERI-BSD virtual machine provided.

Compiling the program to the capability mode produces two capability-related warnings as shown in Listing \ref{lst:Warnings that result from compiling the program into a capability mode}. The first warning indicates that the compiler is confused about how to merge bounds and permissions when adding two capabilities. Resolving this warning requires developers to change or cast the \texttt{offset} variable to a non-capability carrying type, as it only represents an integer and not a pointer, or by switching the order of variables so the capability is driven from the left-hand side variable (the \texttt{buffer}).

The second warning suggests a capability misuse, resulting from casting a provenance-free variable of type \texttt{long} to a file pointer. Solving this warning requires changing variables from \texttt{long} type to a provenance-carrying type (a type that can hold capability metadata) such as \texttt{intptr\_t}. As the \texttt{file} variable is passed through many functions, correctly solving the issue requires developers to change the \texttt{file} variable type in all the functions it is passed in, if they changed it only in the parent function, the error will be gone, but the issue will not be solved.

\code{Warnings that result from compiling the program into a capability mode}{Java}{code/warning.txt}

\code{Code resulting in the first warning}{C}{code/firstissue.c}

\code{Code resulting in the second warning}{C}{code/secondissue.c}

\subsubsection*{Code review task}

Reviewing merge requests is a daily software engineering task. Developers review and test small changes made by individual developers before their code can be integrated with the main code base. Sometimes these changes are new features, but sometimes they are small fixes to address known issues. In this task, developers were told that they have to act as a \emph{merge master} (a software engineer dealing with merge requests) and to review four merge requests made as part of porting programs to the CHERI platform. Participants were asked to answer the following questions for each merge request:

\begin{itemize}
    \item Why do you think this change has been made?
    \item Does this change provide a security benefit; and if so what?
    \item Should you accept this merge request or not?
\end{itemize}

Whilst the porting task focuses on whether the developers can write CHERI code; this task focuses on whether they understand and can tell good code from bad, and understand the reasons why a change has been made. Whilst we were not specifically interested in whether the changes were safe to merge (instead more interested in the developer's perceived rationale for why they were safe to merge), three out of the four merge requests were taken from code changes made to port programs from FreeBSD to CHERI-BSD and should be accepted. The fourth code change includes a function that calculates the pointer size in a way that is not compatible with the CHERI architecture and should be rejected.

\subsection{Analysis}

Our work uses Patnaik's whiffs (Table~\ref{tab:patnaik}) as an initial code book for mapping why developers struggled to use CHERI and go beyond it by adding an additional whiff: \emph{floundering}. Additionally, we coded the interviews looking for signs of struggles and frustration, and we took note of developers' suggestions and thoughts on CHERI. The interview sessions have been through many rounds of coding, and every time a new code emerges, we revise other interviews and add it to the relevant parts of the sessions.

\subsection{Threats to validity}

We have identified the following threats to the validity of this study:
\begin{itemize}
\item The study may not generalize. This study has been conducted with nine participants with C programming experience. We believe that we reached saturation with nine sessions as the same patterns of confusion and floundering were seen consistently, however, a larger sample size that includes developers from different backgrounds could reveal different insights.  We do not claim that the issues we identified apply to all developers that may use CHERI, but they were observed in our study
.
\item  The code developers worked with may not be realistic of how developers work in practice.  We attempted to mitigate this by basing the code on examples and patches from the CHERI BSD project; but the code may still not be representative in small samples.
\item The study was performed in a lab under research conditions. The presence of observers may have influenced participants' performance, leading to the Hawthorne Effect~\cite{jones1992there}.
\item The experience of the authors may have influenced their coding and analysis of the sessions. To help mitigate this, the code book was discussed and reviewed with other researchers to help make any bias more general.
\end{itemize}

\section{Results}

Tables \ref{tab:participation} and \ref{tab:participation2} show the results of the porting and the merge requests tasks. While all of the participants managed to solve the first warning, only six were able to solve the second issue.

\begin{table}[t]
  \caption{Outcomes of the porting task}
  \centering
  \begin{tabular}{l c}
    \toprule
    Warning & Number of participants who fixed it\\
    \midrule
    First warning & 9/9 \\
    Second warning & 6/9 \\
    \bottomrule
  \end{tabular}
  \label{tab:participation}
\end{table}

\begin{table}[t]
  \caption{Outcomes of the merge requests task}
  \centering
  \begin{tabular}{l c}
    \toprule
    Merge request & Participants who decided to accept the merge\\
    \midrule
    First patch & 3/9\\
    Second patch & 4/9\\
    Third patch & 6/9\\
    Fourth patch & 2/9\\
    \bottomrule
  \end{tabular}
  \label{tab:participation2}
\end{table}

\FloatBarrier{}
\subsection{Porting task}
Technically, solving the porting task only requires the participants to change the types of five variables. Whilst all nine of the developers managed to resolve the first warning, only six managed to resolve the second and successfully run the application. The issues highlighted from the task are shown in Table~\ref{tab:codes-porting}. Having identified a series of issues we assigned \emph{usability smells} to each of them by mapping the issues to Patnaik~et~al{.}'s whiffs~\cite{patnaik2019usability}. In cases where no whiff seemed to match, we let a new code emerge or did not map anything.

\subsubsection*{Warning driven}

When attempting to port the code most (7/9) of the developers were \emph{warning-driven}, that is they relied on the compiler's warnings to guide their work without reading much of the code. When faced with a warning, they would look it up in the documentation and search for similar or identical examples, even if some of them already understood the underlying issue and had an idea of how to fix it.

When we asked if the warnings would affect the program at runtime, we received mixed responses: five were confident that the warnings would not affect the running program, and instead, the program would just run without CHERI's protections. It is worth noting that these warnings will actually affect the running program and result in a run-time error.


\begin{quote}
  ``Oh, um\ldots{}well, no, because they are warnings''

  ``I do not think it will affect the program running, because it is only a warning''
  
  ``Well, it is a warning, so probably not''
\end{quote}

The participants seemed to also relate warnings to security, as most of them thought of it as a sign that CHERI protection are not working:

\begin{quote}
    ``It probably means that it is not as secure as it should be or could be.''

    ``It would not have affected the programme working, but it would have affected the benefits that you get from using CHERI''

    ``It will probably break some of the security properties of CHERI if it is not done right''.

    ``The code will work, but you do not have the protections that CHERI has given you if you run it''
\end{quote}

As the developers saw the warnings they showed a good understanding of what could have caused the warnings from the theoretical side, some tried to map these warnings to the bugged code where they could have originated from, while others said they recognized the warnings but were not sure why they are happening in the code, and went straight to the documentation and looked for the exact syntax of the warning.

\begin{quote}
    ``This is just similar thing to the example''
    
    ``I do remember reading about this but I can not remember the reason for it''
    
    ``It has the exact same error code coming out there from the return function, it should be the same thing.''
\end{quote}

\subsubsection*{Good understanding of theory}

The main core of CHERI is capabilities and how they replace the raw pointers in memory. The concept and theory behind CHERI capabilities seemed to be well understood by the participants, as they recognized the ambiguous provenance issue in the first warning, and understood the need for a provenance-carrying type. Participants also manifested their good understanding of other aspects of CHERI such as that capabilities have to be derived from other capabilities, and the implications of having an invalid capability, as that would result in the validity tag being~violated.

\begin{quote}
    ``Both the buffer pointer and the offset pointer have capabilities attached to them, and the problem is that if you are adding them together, they might have different capabilities.''

    ``It does not let the compilers know which of the provenances it should take''

    ``It is ambiguous, but it defaults to using the left-hand side to cast the metadata''

    ``And provenance means, you know, information about where they came from and their origin and their history''
\end{quote}

Despite the participants' understanding of the theory behind CHERI, applying the theory in practice to solve the task was associated with hesitation and a general lack of confidence. Only one participant was able to solve the warnings based on his understanding and without referring to the documentation, but the rest of the participants decided to consult the documentation to look up the warnings and look for identical examples, as they were not very sure how to exactly translate their theoretical understanding to technical solutions.

\begin{quote}
    ``I will search documentation for the source of the problem.''
    
    ``I think this was in the documentation I was looking at earlier''
    
    ``So I remember reading it in the documentation, there was a warning about casting particularly longs''

    ``Cast from provenance for integer type to pointer type. I do remember reading about~this''
    
    ``So, if I just read all of this, I will probably figure out what is going on''
\end{quote}

One participant was not quite sure if these warnings were caused by CHERI or by other bugs, and he decided to compile the program to the base mode, to check if the warnings would still show up, other participants questioned their C programming experience, as they were not sure if some of the data types are CHERI related to not, and whether the changes they are making are actually correct.

\begin{quote}
    ``I’m not experienced enough, probably as a C programmer''
    
    ``Maybe this is my C being rusty''
    
    ``I guess that is my lack of C knowledge coming in''
\end{quote}

The focus of the participants on syntax changes was evident in many sessions, as some participants' leading motive was to find the correct data type to use. The required syntax changes seemed to be unclear to participants especially when solving the second warning.

\begin{quote}
    ``It has the right syntax and based on the CHERI documentation''
    
    ``I just want to double-check exactly what the address type does because it is a new type introduced by CHERI C''
\end{quote}

\subsubsection*{Floundering}

We observed points where developers struggled with the API and seemed \emph{confused}. Developers became confused with how to fix the warnings and CHERI's specific eccentricities.  As one developer put it:

\begin{quote}
  ``A lot of the API documentation kind of just a little bit washed over me.''
\end{quote}

But as we observed the developers attempting to fix the warnings, we noticed a pattern of behavior: developers would make random changes without fully understanding the implications behind these changes. They would make changes in random places or for no clear reasons in the \emph{hope} that it might magically fix things.

\begin{quote}
  ``I'm charging around just hoping to change things!''
\end{quote}

\begin{quote}
  ``I'm pretty sure this will not work, but I want to see what it tells me when I try.''
\end{quote}

\begin{quote}
  ``I just started casting everything to \texttt{uintptr\_t} pointer which I believe will not work.''
\end{quote}

\begin{quote}
  ``Do I need to cast this? Probably. I will just do it because it might work.''
\end{quote}

Despite the developers understanding the basics of CHERI and the underlying concepts behind it, they struggled to solve CHERI-specific warnings and went at it like first-year computer science students in their first programming lab. They tried many solutions and changes at random, hoping that it would make the warning go away. This pattern is different from the \emph{confusion} whiff where developers struggle to use an API or when errors are misaligned with their mental model. The developers understood the underlying cause of the warnings, however, they could not translate this understanding to a practical solution and instead, tried anything they could think of hoping that it \emph{might} work. We created a new smell of this pattern: \emph{floundering}.

\begin{table}[t]
  \caption{Issues emerged when analysing the porting task transcripts. We assigned a usability whiff capturing the underlying issue. Issues are split by the related warning.}
  \centering
  \begin{tabular}{
    p{\dimexpr .1\linewidth - 2\tabcolsep}
    p{\dimexpr .7\linewidth - 2\tabcolsep}
    p{\dimexpr .2\linewidth - 2\tabcolsep}
    }
    \toprule
    Warning & Issue & Usability Whiff \\
    \midrule
    1 & First warning might not be an issue & Confusion \\
    1 & Thought that \texttt{intptr\_t} does not carry capability metadata & Confusion \\
    1 & Changed all types to non-capability types & Floundering \\
    1 & Casted the whole line to \texttt{void} pointer & Floundering \\
    1 & Casted the whole return line to \texttt{intptr\_t} & Floundering\\
    \addlinespace
    2 & Not sure if the warning is CHERI related & Sleuthing \\
    2 & Changed many types and kept running the program to check & Floundering \\
    2 & Developer unsure about the need to change the variable type in all the calling functions & Confusion \\
    2 & Changed types of variables unrelated to the warnings & Floundering\\
    2 & Tried to use \texttt{malloc} to store the string to solve the issue & Floundering \\
    2 & Tried to use signed integer instead of unsigned to fix the issue & Floundering \\
    2 & Trial and error to fix the warning & Floundering \\
    \addlinespace
    Overall & Getting pointer size in CHERI is different than in normal C & Confusion\\
    Overall & Decided to read the code after feeling stuck & Post-mortem\\
    Overall & Changes led to a new warning & Post-mortem \\
    Overall & Developer not sure if changes will fix the code & Floundering \\
    Overall & Copying and pasting from documentation without understanding the meaning behind the changes & Floundering \\
    Overall & Developers did not check if warnings will affect run-time &  \\
    Overall & Not sure which warning is causing the error & Post-mortem \\
    Overall & Searched online for the run-time error as the developer could not find information about it in the documentation & Sleuthing \\
    \bottomrule
  \end{tabular}
  \label{tab:codes-porting}
\end{table}

\paragraph{Ambiguous source of provenance warning}

All participants managed to resolve this warning successfully. Developers understood the issue and they were able to find related examples in the documentation. Solving this warning requires casting the type of one variable to a non-provenance carrying type. Despite the simplicity of the required change, some developers floundered before eventually arriving at the correct fix. Some developers attempted to cast the return statement and variables to various types, one participant cast the whole return statement to a \texttt{void} pointer, and another to \texttt{uintptr\_t}. Others were confused about whether to flip variables or change types in order to fix the warning. Only one developer solved this issue on the first try.

\paragraph{Capability misuse warning} 

To fix this warning developers had to change a \texttt{long} variable to a capability-carrying type in all the calling functions. However, if developers only change the variable type in the top-level function, the warning would disappear but the issue will persist and a security exception would crash the program when running it.
Out of the nine participants, only two realized at the beginning that the change is needed to be propagated to other functions. Even so, these developers were not entirely sure if they were doing it correctly and were not confident about their solutions. 

\begin{quote}
  ``I'm wondering whether I solved the problem for the first run of things''
\end{quote}  
\begin{quote}
  ``Could be digging myself into a hole, but let us carry on''
\end{quote}  

In other sessions, developers seemed \emph{confused} when the warning disappeared, but the issue persisted. Once the developers encountered the security exception, they began to show \emph{floundering} signs, attempting fixes at random. Only one developer attempted to debug the run-time error using a debugger (GDB), but did not arrive at the cause of the error. Part of what caused floundering and confusion was that the developers could not find any help in CHERI's documentation for the run-time exception:

\begin{quote}
  \emph{Interviewer}: ``What are you looking for?''

  \emph{Developer}: ``Basically a fix to the security exception, I do not know where to look.''
\end{quote}

As part of floundering, some decided to replace some pointers in the code with \texttt{char} pointers since they had seen them used in the documentation. Others attempted to use \texttt{malloc} to store the string value that is written to the file, and some cast a lot of variables to pointers. None of these changes were correct or needed to solve the warning. Developers made changes and then ran the program to test if their changes would work or not. Again, participants showed a lack of confidence in their solutions, and whilst some ended up with a working solution, extra unnecessary changes were also made as one participant changed every variable of the type \texttt{long} to \texttt{uintptr\_t} hoping for things to work. Three developers eventually gave up. They understood the theory of what was happening and why the warning was happening, and they understood the concept of provenance types and capabilities, but struggled to apply their knowledge to fix the warning, and basically \emph{floundered}.

\subsection{Pull request task}

When completing the pull request task, the developers were asked to decide whether a patch should be merged or not and whether the merging of the patch had security implications. Our codebook seems to indicate fewer usability smells and instead more general confusion about the CHERI platform itself.  Whilst there is evidence for \emph{floundering} and \emph{confusion}; most of the codes seem to represent developers being unsure, confused about certain types, and struggling with a difficult documentation. Besides questioning whether they \emph{should use something} or \emph{how to use it} (signs of the confusion whiff), some participants seemed cautious and more generally unclear. The codes captured when analyzing developers' narrations while doing this task are shown in~Table~\ref{tab:codes-pull}.


\subsubsection*{MR1}

This merge request was taken from \texttt{kldstat} program in FreeBSD. The \texttt{vaddr\_t} type represents a virtual address, and it does not have the capability information associated with it. We were interested in participants' thoughts about this change, more than their actual decision since they did not have full access to the source code of the program and were not aware of the context of the change.

\begin{lstlisting}[language=diff]
- #define PTR_WIDTH ((int)(sizeof(void *) * 2 + 2))
+ #define PTR_WIDTH ((int)(sizeof(vaddr_t) * 2 + 2))
\end{lstlisting}

Developers showed signs of \emph{confusion} when reviewing the merge request that used \texttt{vaddr\_t} and found it difficult to explain when that type should be used, despite showing that they understood CHERI concepts such as provenance carrying types and capabilities. They were able to find references for \texttt{vaddr\_t} in the documentation, but still could not understand where and how it should be used in general. Some thought that this change needed to use \texttt{alignof} instead of \texttt{sizeof}, others were not sure if it was necessary or not, and others thought that the change was important as it uses a \emph{CHERI data type}.

\begin{quote}
    ``I still can not confirm if it is a necessary change.''

    ``I do not know if that is a useful change''

    ``I do not think they are using the appropriate variable here to represent the size of a pointer address or size of a pointer in CHERI.''

    ``Without this I do not think the code would work at all''

    ``That still means things will either throw warnings or errors when you are trying to use it''

    ``The fact that it uses size of, instead of align of, which is a requirement in the CHERI architecture basically''

    ``I will reject on the grounds of making what it does less readable''

    ``I think I would reject it, it needs to be align of, but it is not clear from the documentation what it should be''
\end{quote}

\subsubsection*{MR2}

The second merge request was taken from \texttt{ipfstat} program, and it uses \texttt{uintptr\_t} data type to represent pointers to list items. The \texttt{n\_value} field is of a \texttt{long} type (participants were told that) and the program converts that variable into a void pointer, as it holds an address. Under normal C, as the \texttt{n\_value} contains an address, converting it to a void pointer is allowed and that pointer can be de-referenced later. With CHERI, converting longs (a provenance-free type) to pointers will lead to a pointer without the capability metadata and an error will be thrown when attempting to de-reference that pointer. As this pointer has to hold capability information, a type like \texttt{uintptr\_t} must be used to maintain the capability~information.

\begin{lstlisting}[language=diff]
- ipsstp->iss_table = (void *)deadlist[18].n_value;
- ipsstp->iss_list = (void *)deadlist[17].n_value;
+ ipsstp->iss_table = (void *)(uintptr_t)deadlist[18].n_value;
+ ipsstp->iss_list = (void *)(uintptr_t)deadlist[17].n_value;

- frauthstp->fas_faelist = (void *)deadlist[1].n_value;
+ frauthstp->fas_faelist = (void *)(uintptr_t)deadlist[1].n_value;
\end{lstlisting}

The participants' reactions to this change were mixed, some thought this change was wrong, as it does not make any sense to them for the \texttt{n\_value} to be stored in a capability type, while others were confused about what would happen if a \texttt{long} value is converted to a capability. Some thought that this added to the security of the program since it is \emph{CHERI complaint}, and others thought that this change was only made to silence a warning somewhere.

\begin{quote}
    ``That represents an address, and you want to turn it into a CHERI capability, but that is a bit useless because it is just a \texttt{long} value''


    ``It feels completely unnecessary''

    ``They are doing this to silence a warning somewhere''

    ``Change must be made, because we are giving a long, which is not capability''

    ``This is a necessary change to make it CHERI-compatible to preserve the capabilities''
\end{quote}

\subsubsection*{MR3}

This change is taken from \texttt{crypt\_server} software when ported to CHERI-BSD. The older version of the software contained a legacy way of defining functions in C (Unprototyped (K\&R) functions).
While this is not a problem for any code written in the recent years, we were interested in what ways developers would associate this change with CHERI. CHERI does not allow this legacy way of defining functions as pointers have to be stored in capability-aware variables.

\begin{lstlisting}[language=diff, breaklines=true]
- int _arcfour_crypt(buf, len, desp)
- char *buf;
- int len;
- struct desparams *desp;
+ int _arcfour_crypt(char *buf, int len, struct desparams *desp)

- _my_crypt = (int (*)())dlsym(dlhandle, "_des_crypt");
+ _my_crypt = (int (*)(char *, int, 
+ struct desparams *))dlsym(dlhandle, "_des_crypt");
\end{lstlisting}

Six developers decided to allow this merge request, some thought this change was not related to CHERI and that without it, the program would not compile, others thought it was to make the program more compatible with CHERI, as discarding data types would invalidate the capability metadata. Two participants said they were not sure whether to accept or reject the request, and one participant rejected it on the basis that they could not justify it.

\begin{quote}
    ``I would say the change makes it read better. Because you have the parameters specified in in this, so I guess it is a little bit less ambiguous. But I'm not sure if it is doing CHERI stuff''

    ``I’m not so sure that I have seen that in the documentation''

    ``That is not going to work because you are not going to have any of the metadata set''

    ``It is supposed to prevent things that CHERI prevents''

    ``There does not seem to be anything obviously different for CHERI''
\end{quote}

\subsubsection*{MR4}

This merge request replaces a static pointer size with a function that infers the pointer size based on the size of a void pointer. While this can work on normal computers, it would not work with CHERI-supported hardware, as pointer size is not only the size of the address but rather, the size of a CHERI capability, as pointers are not only addresses in CHERI.

This merge request should be rejected as it does not account for the capability metadata. When reviewing the merge request, developers struggled generally with the concept of how to get the size of a CHERI pointer, and whether the \texttt{sizeof} function will yield the address size or the capability size.

\begin{lstlisting}[language=diff]
- int pointer_size = 32;
+ int pointer_size = get_pointer_size_in_bits();

+ static int get_pointer_size_in_bits(){
+    // size in bytes
+    int size = sizeof(void*);
+    // check if pointer size is 32 bit or 64 bits
+    if(size*8==32){
+        printf("%s","This program supports 32 bit machines");
+        return 32;
+    }else{
+        printf("%s","This program supports 64 bits machines"); 
+        return 64;
+    }
+ }
\end{lstlisting}

Seven participants rejected this merge request, some thought it would lead to bad pointer calculation, others said that it would lead to the program being broken, and others said that it would break the security features of CHERI as it would set incorrect bounds. The two participants who accepted this merge request did not notice that CHERI pointers were larger in size.

\begin{quote}
    ``The security implication would be getting that wrong and writing over stuff.''

    ``I'm reasonably certain this  can not be merged because again we are forgetting the extra space of provenance''

\end{quote}

One participant thought that \texttt{sizeof} should not be used to fetch the pointer size, but instead \texttt{alignof}, as he saw it in the documentation. The participant is not wrong as the documentation actually advises to use \texttt{alignof} instead of \texttt{sizeof} in a footnote, but that is specific to the case of allocators that try to use \texttt{sizeof} to align their return values, CHERI advises to use \texttt{alignof} in that specific case, but not as a replacement of \texttt{sizeof}.

\begin{quote}
    `` I’m not entirely sure whether good practice is \texttt{sizeof} instead of \texttt{alignof}''
\end{quote}

\subsubsection*{CHERI means more secure}

We noticed that sometimes, in both the coding and reviewing tasks, the participants associated the use of CHERI types with enhanced security and the correctness of the solutions.  This suggests that whilst they correctly associate CHERI with increased programming safety, the reasons why it does that are not quite as clear and somewhat magical to developers, in many occasions in both tasks, the participants associate CHERI types with the correctness of the implementations, regardless of the reason.

\begin{quote}
\end{quote}  
\begin{quote}
  ``I'm not so sure security wise, but it is definitely good practice to have it explicitly cast it to that sort of pointer \verb|uintptr_t|''
\end{quote}  
\begin{quote}
  ``Probably accept because it looks better and if it does make something comply with CHERI standards, then that is probably better, right?''
\end{quote}  
\begin{quote}
    ``Because it has the right syntax and based on the CHERI documentation''
\end{quote}

\begin{table}[t]
  \caption{Issues highlighted when analyzing the merge-requests task and their mappings to usability whiffs.}
  \centering
  \begin{tabular}{
    p{\dimexpr .1\linewidth - 2\tabcolsep}
    p{\dimexpr .7\linewidth - 2\tabcolsep}
    p{\dimexpr .2\linewidth - 2\tabcolsep}
    }
    \toprule
    Task & Issue & Usability Whiff \\
    \midrule
MR1 & It is not clear from the documentation how to solve this & Sleuthing \\
MR1 & Confused how CHERI handles the size of pointers & Confusion \\
MR1 & Can not confirm if it is a necessary change & Confusion \\
MR1 & Mixed between \texttt{vaddr\_t} and a pointer & Confusion\\
MR1 & Not sure about the sizes of \texttt{vaddr\_t} and pointers & Sleuthing\\
MR1 & Unclear if it has security implications & Confusion\\
      \addlinespace
MR2 & Confused about security implications & Confusion \\
MR2 & It does not make sense to cast \verb|int| to \verb|void| & Confusion \\
    \addlinespace
MR3 & It does not have anything obvious with CHERI & Confusion \\
MR3 & Not sure if it is CHERI related & Confusion \\
    \addlinespace
MR4 & Reject because of strange implementation & Confusion \\
MR4 & The change might be unnecessary & Confusion \\
    \bottomrule
  \end{tabular}
  \label{tab:codes-pull}
\end{table}

\FloatBarrier{}

\section{Discussion}
\subsection{Something's fishy\ldots}

During several sessions we noticed the following when developers worked with CHERI code: When developers are presented with an error message or code that they did not immediately fully understand how to fix, they would start searching the documentation for the literal pattern. When they found something similar enough they would attempt to use the documentation to fix the error, but when their fix failed to work or left questions in their mind they became slightly panicked and would attempt solutions somewhat at random in order to fix the issue. As two developers put it:

\begin{quote}
   ``I might have to do a bit of trial and error here\ldots''
\end{quote}
\begin{quote}
   ``I'm charging around just hoping to change things.''
\end{quote}

Random solutions seemed to include switching arrays for \texttt{malloc}'d pointers and making everything pointers (because CHERI is understood as having a different pointer architecture) and making random changes in the hope that something might fix the error. The participants would blame either the code for being fundamentally incompatible with CHERI, or themselves for not being clever enough to understand it.

The floundering whiff is concerning, unlike the \emph{confusion} whiff which suggests that developers are struggling to align their own mental models with CHERI's, the floundering whiff indicates that despite developers' understanding of the theory of CHERI and the concept of CHERI capability and provenance, working with such concepts in practice is not easy. They floundered while attempting to fix the code by randomly trying things that \emph{might work} just in case they can make the errors go away.

Our study highlights that there are usability issues with CHERI lurking. The \emph{floundering} smell was seen repeatedly throughout the study reducing developers to guessing and hoping that their programs were correct and secure. Not digital security by design; and a sign that CHERI needs a better explanation for most developers. The reasons behind this whiff seem to stem mainly from the complex documentation and the display of warnings and errors. 

Yet there are grounds for hope that as familiarity with CHERI and capability-based hardware increases developers will flounder less.  As part of the study, we asked participants to describe what a capability was before taking part. We repeated the same question again afterwards to look for signs of learning effects. Participants' answers after the task generally showed more specificity and detail about the changes needed to port programs to CHERI.  This suggests that as developers work more with CHERI, familiarity may help them better understand what they need to do and \emph{perhaps} flounder~less.

\subsection{What does it mean for CHERI to be usable?}

Some of the issues found in our study might not be conclusive to CHERI, for instance, a poorly documented API will always result in developers struggling, regardless of the platform, however, some of the issues we found highlight areas where CHERI can be improved further to be easier to learn and use. The floundering smell discussed above is a dangerous sign for a platform that aims for security, especially since many of the participants implemented either incorrect or unnecessary changes in the code, just to make the warning go away. In a tiny codebase like the one in the study, this might not be an issue, but for a larger codebase, implementing such changes could result in breaking the functionality of the program, or creating more bugs/warnings. 

For CHERI to be usable, developers must be assured in their answers. The act of frequent look-ups to the documentation is not a usability issue in itself, but rather how accessible the answer is, and CHERI documentation was challenging for developers as they had to search long pages that contained much information they did not really need at the time of fixing the task.

While CHERI is not a programming language, it has an effect on the compilers running on top of its hardware, thus new data types were introduced to C/C++ and modifications were made to their compilers. Our study looks at how CHERI is received from the developer end, and our participants did not only have to learn new C semantics but also had to read and understand the theory behind CHERI. While many of the issues the participants faced were related to how the C compiler treats warnings and errors, these issues were fundamentally related to CHERI as, first, they were caused by new changes introduced by CHERI, and second, to be fixed, the participants had to understand many new concepts and ideas related to CHERI.

\subsection{Scope for improvement}

\begin{table}
  \centering
   \caption{Codes capturing suggestions participants made for improvements needed to make working with CHERI easier; grouped by whether they were part of the compiler tooling or the documentation.} %
  \begin{tabular}{ll}
    \toprule
    Code & Category \\
    \midrule
    Better way to get pointer size & Compiler\\
    CHERI compilers should help the developer spot these problems & Compiler\\
    Compilers should specifically say what types need to be fixed and changed & Compiler\\
    Compilers should tell developers where in the code the changes have to be made & Compiler \\
    \addlinespace
    Documentation should show more examples that has context to them & Docs\\
    Less writing and more simple examples & Docs \\
    Need for a cheat-sheet & Docs\\
    Need for an easy conceptual model & Docs \\
    Need for less verbose documentation & Docs\\
    Simpler documentation of needed changes & Docs\\
    Tutorial with a simple program then introduce changes to it & Docs\\
    \bottomrule
  \end{tabular}
  \label{tab:improvements}
\end{table}

When completing the tasks, participants sometimes remarked on the usability challenges they were facing and what they felt could be done to improve them. We created codes to capture their suggestions. Broadly speaking, these usability issues are related to either the \emph{compiler} or the \emph{documentation} as shown in Table~\ref{tab:improvements}.

\paragraph{The compiler.}
Programmers are well known to ignore warnings \cite{gorski2020listen}, yet when dealing with a new architecture, this behavior can be especially dangerous. Several of our participants seemed confused about whether it was safe to ignore CHERI-specific warnings, and yet in every case, it was not. The warnings were identifying an underlying issue, and some participants assumed that the warnings would not affect the program when running it:

\begin{quote}
``I need to double check whether this warning is relevant or not.''
\end{quote}
\begin{quote}
\emph{Interviewer}: ``So do you think these warnings generally will affect the programme if you run it?'' \\
\emph{Developer}: ``Oh, um\ldots{}well, no, because they are warnings.''
\end{quote}
\begin{quote}
``I do not think it will affect the program running, because it is only a warning.''
\end{quote}
\begin{quote}
``Well, yeah, I mean given it is a warning, it may not be a change that I need to do anything about.''
\end{quote}

Given that these warnings would lead to errors, perhaps it needs to be made more explicit to developers that these are issues that need fixing and should be treated as \emph{errors} instead.

Additionally, since the compiler knows what it is compiling, additional warnings about particular concepts that are likely to catch developers out when working with CHERI C (for example, surrounding the use of capabilities) could help developers spot issues ahead of time. Using empirically validated techniques to help ensure that the compiler output is readable may also help \cite{denny2021designing}.

Furthermore, one of the issues that developers struggled with the most is that in some cases, the warnings will be gone, but the issue will still be there, and this happens because the developers made the correct change in one place, but that change is needed in other places and in the calling functions as well. The compiler should be fixed to keep the warning until the correct change is implemented in all the needed places in the code.
    
\paragraph{The documentation.}

Whilst CHERI is still new, expecting the documentation to be polished at this early stage is perhaps unfair, but it is clearly an area programmers struggle with:

\begin{quote}
``It is not fully clear what to look for. I have looked for these two types in here and I can not really spot the main reason to why I would change it.''
\end{quote}
\begin{quote}
``It is just very poor documentation, so it has not really got a basis actually. There is no simple hello-word example of how you can just use the pointers and other things.''
\end{quote}
\begin{quote}
  ``Just taking the documentation as it is and taking the recommendation, that did not work unfortunately.''
\end{quote}
\begin{quote}
``I think the thing lacked any sort of tutorial. It jumped straight into fairly hardcore descriptions of it all, and there is no warm up.''
\end{quote}
\begin{quote}
``Yeah, there is very few examples in here of past single lines extracted from programs, rather than more context in there.''
\end{quote}

Lack of tutorials, lack of CHERI examples, and simple errors led to programmers becoming frustrated. Other participants expressed how broad the documentation was and just wanted simpler examples and cheat-sheets: providing more focused documentation for developers who just want to work with CHERI without understanding its full features and security architecture.
Whilst we can not expect all the tutorial material to be there before the widespread adoption of digital security by design architectures---the calls for it will only get louder.

\paragraph{Document run-time errors.}

Additionally, when participants ignored the warnings or when the warning disappeared but the issue was not fixed, it typically led to runtime errors. Developers struggled to identify these errors and were just left in a state where their programs were crashing without fully understanding why. They assumed their programs would work because they had either quashed the warning or followed the guidance in the documentation that had not resolved it and seemed surprised that bugs persisted. As two developers said:

\begin{quote}
  \emph{Interviewer}: ``Why do you think this will work?'' \\
  \emph{Developer}: ``Because it says it will.''
\end{quote}
\begin{quote}
  ``Well, the warning was solved, maybe the problem was not solved.''
\end{quote}

The documentation does not have any mention of run-time errors.
Ensuring that when CHERI safety properties are violated leads to visible and explicit errors that developers can quickly understand would help developers know and better understand why things are failing if and when they do. 

\section{Conclusion}

As one developer put it:
\begin{quote}
``CHERI stops your code from violating memory and security, but it doesn't stop the programmer from creating bad code.''
\end{quote}
The security possibilities of digital security by design offer a great opportunity to improve the safety of all software and hardware at a fundamental level.  But early results from our study suggest that we are at risk of falling into the same usability pitfalls that plague conventional hardware: namely, poor documentation, and confusion about warnings and errors.  These are all well known usability smells~\cite{patnaik2019usability} for programmers, but to see them appearing with CHERI hardware suggests that there are still improvements to be made.

No matter what clever hardware we have, we \emph{cannot} ignore the human element and the programmers building on these systems.  It doesn't matter if we have secure by design hardware if we cannot build the software for it.  Currently developers seem to get confused and flounder when using CHERI.  If we're going to get widespread adoption then it will help if adopting CHERI is \emph{a piece of cake.}

\section{Acknowledgements}
This work was funded until 2024 by the UKRI Digital Security by Design (DSbD) Programme, to support the DSbD ecosystem.

\FloatBarrier{}
\bibliographystyle{splncs04}
\bibliography{bib}

\appendix
\section{Task code}

\code{CHERI study}{C}{code/cheristudy.c}

\end{document}